# Conceptualizing Business Process Maps

Geert Poels[1], Félix García[2], Francisco Ruiz[2], Mario Piattini[2]

[1] Faculty of Economics and Business Administration,
Ghent University
Belgium
geert.poels@ugent.be,

[2] Institute of Information Technology and Systems
University of Castilla-La Mancha
Spain
{felix.garcia, francisco.ruizg, mario.piattini}@uclm.es

**Abstract.** Process maps provide a high-level overview of an organization's business processes. While used for many years in different shapes and forms, there is little shared understanding of the concept and its relationship to enterprise architecture. In this report, we position the concept of business process map within the domain of enterprise architecture. Based on literature, we provide a conceptualization of the process map as a business process architecture model that can be integrated with the broader enterprise architecture model. From our conceptualization we derive requirements for designing a meta-model of a modelling language for process maps. The design of this meta-model is the subject of a research paper, entitled Architecting Business Process Maps, for which this report acts as a complement that details the underlying process map conceptualization.

## 1. Introduction

Business Process Architecture (BPA) has recently received attention in Business Process Management (BPM) research [1]. BPA refers to the fundamental organization of an enterprise's system of business processes. In the Enterprise Architecture (EA) field, the *business process architecture* is considered part of the business layer of an organization's enterprise architecture, while in the BPM field it is presented as an artefact that results from the BPM lifecycle process identification phase during which the organisation's business processes are identified and prioritised [2]. Whereas the purpose of the business process architecture in BPM is to single out the processes that will be subjected to further BPM lifecycle activities, in EA the business process architecture is essential for analysing and improving enterprise coherence, enterprise agility, business/IT alignment and strategic fit [3, 4].

In BPM, a model describing (part of) the business process architecture is commonly, though not uniquely referred to as a *process map* [5]. The concept of process map as a holistic and abstract representation of an organisation's interrelated business processes, has only recently been investigated [6], while being in use for many years in different shapes and forms. In practice, there is little shared understanding of the concept, related to its contents, form and purpose. According to [7], the current variety in process maps that can be observed might be due to the lack of modelling language dedicated to expressing process maps. The need for designing such language, preferably building upon a general-purpose modelling language, has been expressed by different researchers [1, 8, 9].

Apart from the lack of specific modelling language for process maps, there is unclarity in the conceptualisation of the process map in relation to EA. In BPM, process maps are not integrated in broader enterprise architecture descriptions. Consequently, they cannot be exploited at full scale for aligning BPM initiatives with an organization's architectural vision.

The goal of the research presented in this report is to provide a conceptualization of process maps in the context of EA by considering a modelling language for process maps as an architecture description language. As a result, we conceptualize the process map as an enterprise architecture artefact and present requirements for a business process architecture description language that can be used to represent process maps. The design of a meta-model for such language is the subject of a research paper, entitled Architecting Business Process Maps, for which this report acts as complement that details the underlying process map conceptualization.

The remainder of this report is structured as follows: Section 2 presents the methodology we used for conceptualizing process maps. Sections 3 to 8 present the concepts deemed relevant for concept maps. Finally, section 9 presents the conclusion.

## 2. Methodology

We performed a literature review of research on modelling of business process architecture. Research presenting requirements for process maps, informal solutions, and reviews of design approaches for business process architecture was analysed. The result was an inference of different perspectives of process maps in relation to business process architecture, indicating basically a lack of 'architectural point of view' in BPA research, which motivated our research. The primary basis for the conceptual analysis of the process map concept was the process map meta-model of Malinova and Mendling [6], which is to the best of our knowledge unique in its kind. The proposed meta-model is positioned by its designers as a model of a process map conceptualization rather than a formal meta-model defining a modelling language for process maps, which makes it a valuable starting point for our conceptual analysis. Its foundation in BPM research results, however, in a number of assumptions related to the use of process maps prior to BPM implementation (i.e., process identification) and during BPM implementation (i.e., process model management). Specifically, the meta-model conceptualizes the process map as an entry-level model of a hierarchically structured business process *models* architecture instead of a model of the business process architecture. We therefore build our conceptual analysis also on other related work that, although less elaborate than the work of Malinova and Mendling, positions the process map as an abstraction of the business process architecture rather than as part of a hierarchical structure of business process models.

We support our literature-based analysis through the use of *concept maps*, a well-known tool for expressing and sharing knowledge, using the free CmapTools knowledge modelling kit (https://cmap.ihmc.us/cmaptools/).

## 3. Process as Atomic Element

The central concept of the process map is the *Process*. Consistent with most of the reviewed literature, we explicitly chose here for the process map as a black-box model of (part of) the system of business processes of an organization that does not show interior details of these processes. Note that this makes the process map concept different from its definition in the BPM textbook of Dumas et al. [2] as they consider it an abstract process model showing main activities and sequence flows, whereas the model that depicts processes as atomic elements is what they call the process landscape model, which is positioned at the top of their hierarchically structured business process models architecture. To further avoid such terminology confusion, we propose the following *solution requirement* consistent with our research goal of designing a general representation language for process maps within the context of our enterprise architecture description conceptual frame of reference:

***Req. 1: The business process is the atomic element of the process map.***

We further motivate this requirement by the existence of business process modelling languages (e.g., BPMN), which can be used to model business processes at more fine-grained levels of modelling abstraction [10]. Hence, there is no need for a modelling language for process maps to model such finer-grained elements like tasks and gateways. Following Van Nuffel and De Backer [9], the process map should thus only be a black-box model of the organizational processes.

## 4. Inputs, Outputs, Resources, Actors, and Roles

Besides from processes, a process map may show other things and events related to the processes. Processes are triggered by inputs and yield outputs. An *Input* is a request to do something (i.e., start process execution) that can be accompanied by data or physical objects. An *Output* is a tangible product flowing out the process or an intangible result that is the effect of the process execution, in which case the output is commonly referred to as an outcome (e.g., staff trained as a result of a training process, deliverable validated as a result of a validation process) [11]. Such inputs, outputs and outcomes might be shown in a process map [7], though this is not common as it depends on the desired level of abstraction at which the business process architecture is described.

A business process included in a process map can use one or more resources and can have one or more actors responsible for it, which again can be shown or not shown (depending on the desired level of abstraction). A *Resource* refers to any means used during process execution to transform inputs into outputs. An *Actor* is defined as being responsible for the process, e.g., someone in the process manager role [6]. For instance, Van Nuffel and De Backer [9] distinguish a process owner view of the business process architecture in which the process map clearly identifies process ownership.

In other examples of process maps also organizational units or business functions are identified as being involved in process execution, see e.g., [1, 2, 5]. Physical persons or organizational units (e.g., departments, project teams) having certain functions and responsibilities in the organization are typically included in enterprise architecture models, hence linking them explicitly with processes according to specific roles towards business processes (i.e., as actor or resource) connects or even integrates business process architecture models with(in) enterprise architecture models.

Figure 1 summarizes the relationships between business processes and the concepts discussed in this section.

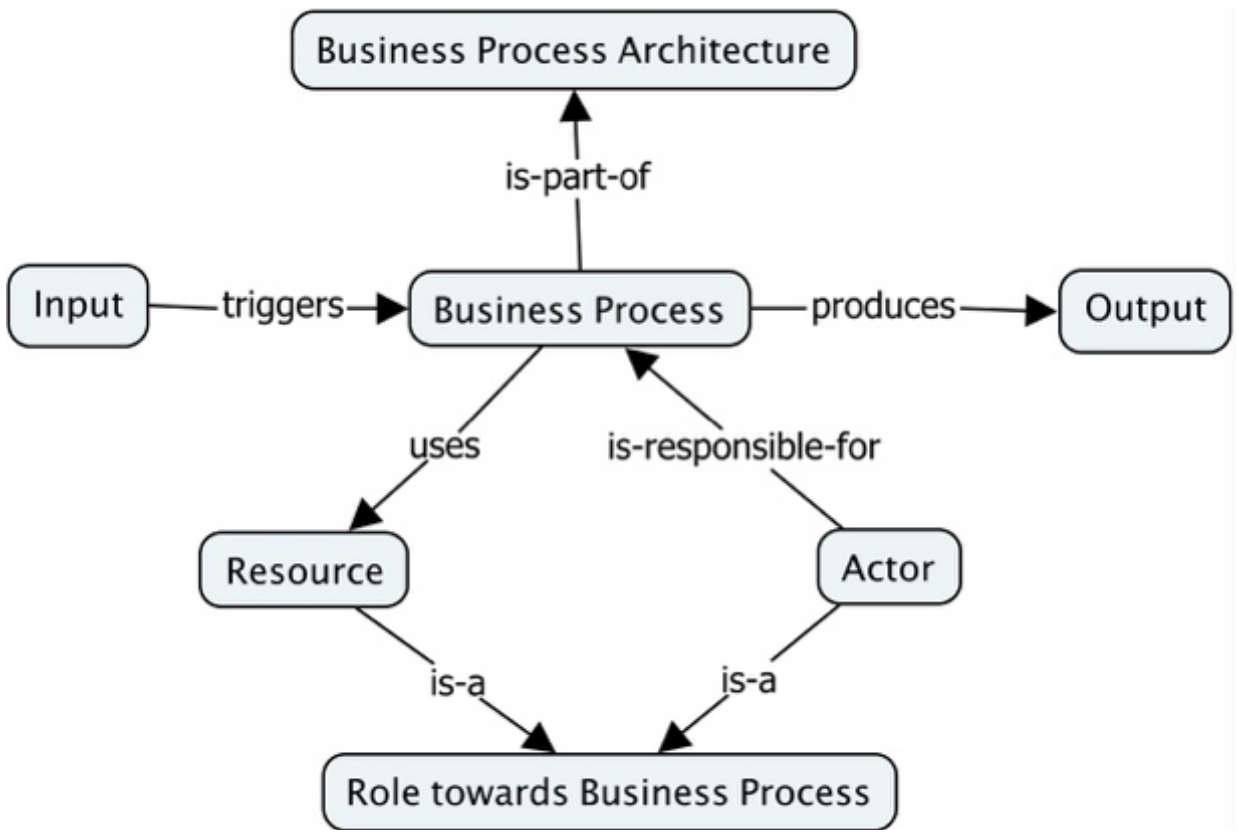

**Fig. 1.** Business processes and their inputs, outputs, resources, actors and roles.

## 5. Triggers and Customers

Figure 2 shows a concept map about how customers trigger business processes through the inputs they provide. A *Customer* is the beneficiary of the business process, which means that the output of the process is somehow of benefit to the customer.

Based on the meta-model of Malinova and Mendling [6], the concept map distinguishes internally and externally triggered business processes. It is shown that business processes can be triggered either by external customers, meaning customers outside the organization, or by internal customers which can be employees (in a certain role towards the business process) or other processes of the organization. Externally triggered processes produce output for external customers, whereas internally triggered processes produce output that is used by internal customers.

Externally triggered processes can act as internal customer for an internally triggered process, which then produces internally consumed outputs. To eventually get outputs delivered to external customers, some decomposition relation between externally triggered processes and internally triggered processes is needed (not shown in Figure 2). We come back to relations between business processes in section 8.

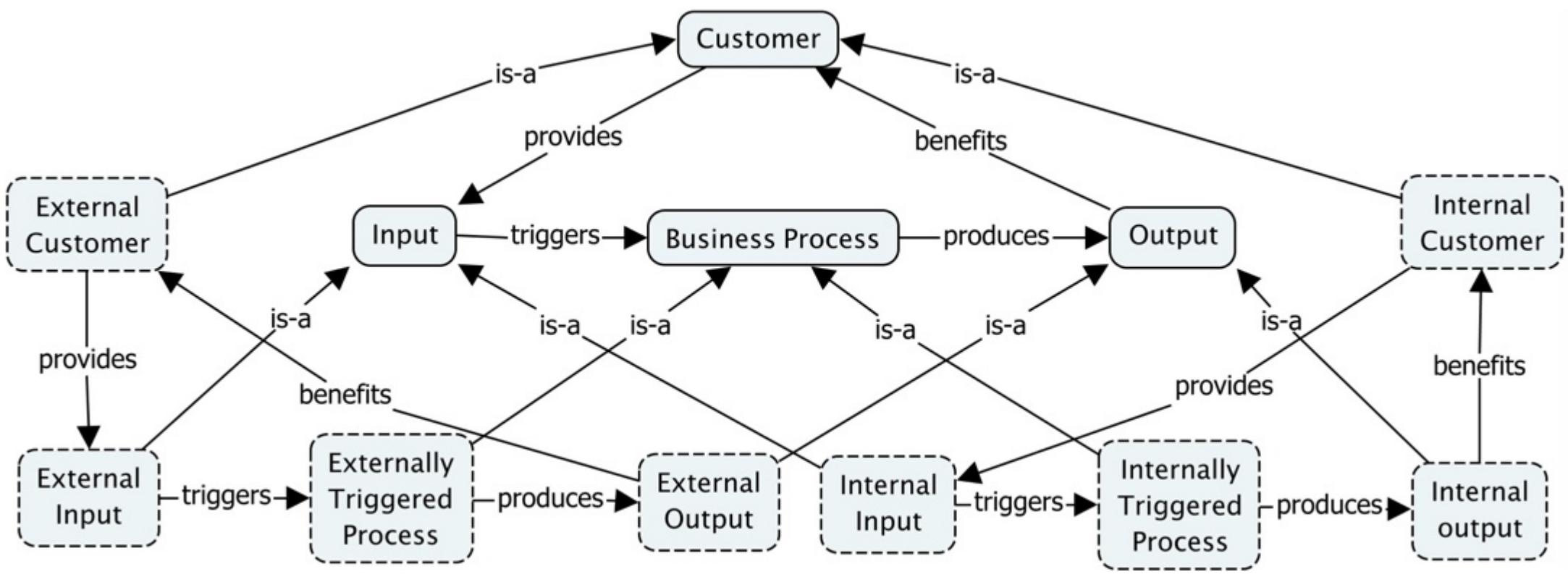

**Fig. 2.** Differentiating internally and externally triggered business processes.

Representing customers of process outputs on process maps is not uncommon, see e.g., [5, 8], though it depends again on the desired abstraction level at which the business process architecture is described, which is, apart from [9], not explicitly defined in the BPA literature that we reviewed.

## 6. Services and Objects

Other relevant process-related concepts found in process maps are service and object. While service is not defined in the reviewed BPA literature, in Service Science, service is defined as an act provided by a party to a beneficiary [12]. In the context of BPA, we accordingly define *Service* as the benefits experienced by the customer of a business process. These benefits inhere in the outputs produced by the process (see Figure 3). While Malinova and Mendling [6] consider only employees as beneficiaries, it is not uncommon to define provided services as an essential property of business process architecture, regardless whether the service beneficiaries are internal or external to the organization [1].

In the reviewed BPA literature, it is very common to observe in business process architecture models the objects that are affected by business processes [1, 5, 9]. An *Object* represented in a process map can have a variety of meanings. Dijkman, Vanderfeesten and Reijers [1] even identify a separate object-based class of business process architecture design approaches (i.e., architecture frameworks) that provide guidelines to derive a business process architecture from a business object model (e.g., the RIVA process architecture framework [13] or the fractal enterprise model [8]). Dijkman, Vanderfeesten and Reijers [1] also distinguish between permanent objects (e.g., a client that is being helped) and case objects (e.g., an order that is being processes), and the reference in [6] to a process flow unit seems to refer to the latter interpretation, although process flow units may also include abstract objects (e.g., product quality). Clearly, whatever interpretation is used, objects are closely linked to inputs and outputs or even to both (e.g., the object is input to the process which changes or transforms it, adding value to it, and returns it again as output, where the value added is the service provided by the process to the customer).

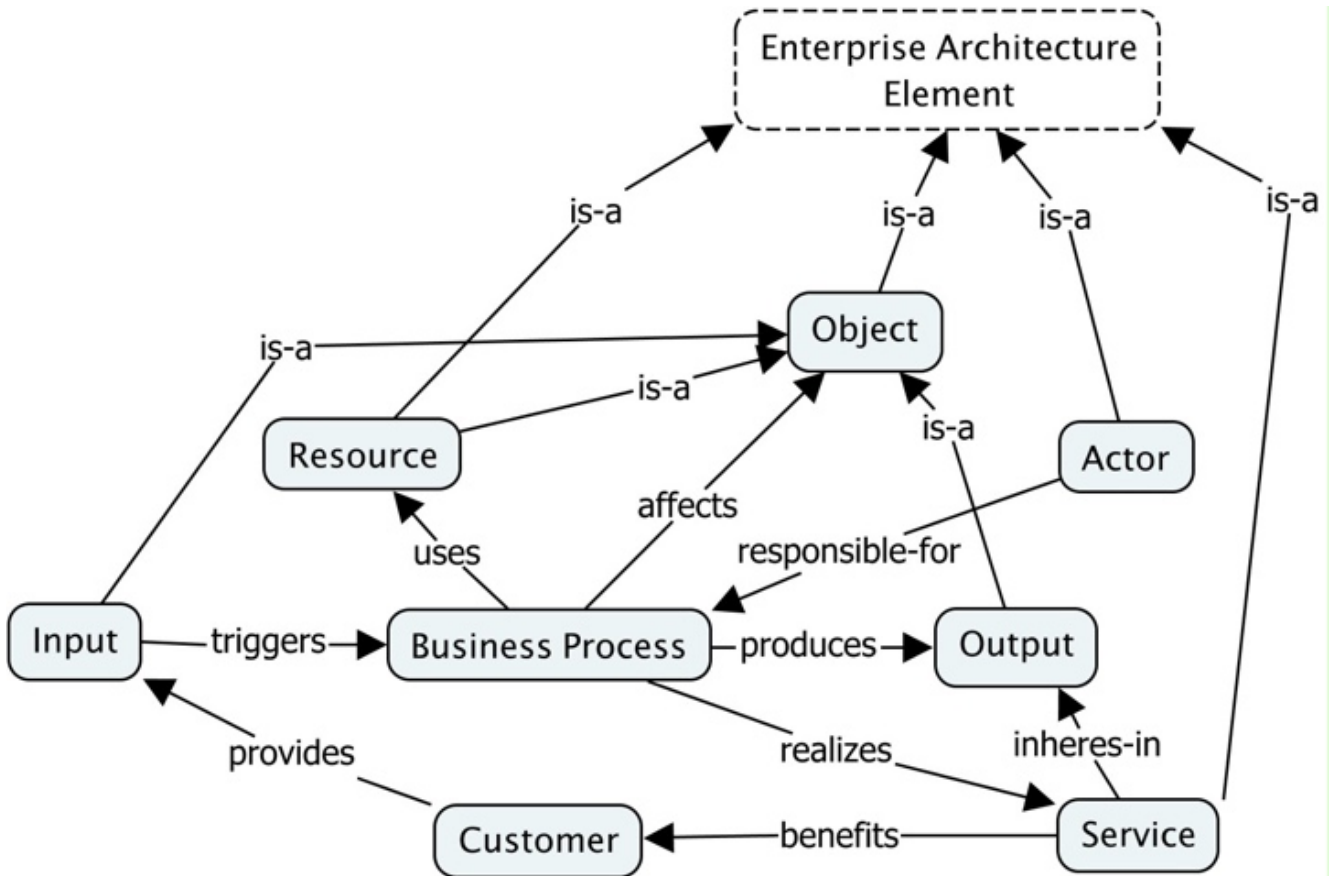

**Fig. 3.** Services, objects, resources and actors as enterprise architecture elements.

Resources used or consumed in business processes (e.g., general supplies like computers and office materials) can be objects of other processes (e.g., the general supplies purchasing, management, maintenance, security, etc., processes) [8], so resources are objects in a particular role of being used or consumed.

It is important to note at this point that actors, objects (and their roles in processes like input, output, resource, etc.) and services can all be represented in enterprise architecture models, so their inclusion in a process map allows positioning the business process architecture within the broader enterprise architecture (see Figure 3). Hence, we notice on the one hand a need for representing in a process map explicit linkages between business processes (and their composites, confer infra) and relevant enterprise architecture elements. On the other hand, what is relevant is largely determined by the purpose served by business process architecture, which is seldom explicated. Therefore, we propose a second solution requirement consistent with our research goal of designing a general representation language for process maps within the context of our architecture description conceptual frame of reference:

***Req. 2: It should be possible to show on a process map the enterprise architecture elements that a business process (composite) is related to.***

## 7. Business Process Groups

One essential element of business processes from an architectural point of view is how they can be grouped. A *Category* is a group of business processes that fulfil a similar role in the organization or serve a similar purpose. Malinova, Leopold and Mendling [7] present an empirical analysis of 67 process maps which indicated that more than 90% categorize the organizational processes according to Porter's Value Chain framework [14], which seems to be a major business process architecture framework. Porter [14] distinguishes between core (or value creation), support and management processes. Sometimes additional categories are added like analyse & measure processes. Other categorizations exist. The RIVA process architecture framework has case processes, case management processes and case strategy processes [13]. Rummler and Ramias [15] make a distinction between launched, sold and delivered processes. Snowdon and Kawalek [16] make use of five categories: operational processes, coordinating processes, controlling processes, monitoring processes, and strategy processes. Depending on the categorization taxonomy used, processes may belong to more than one category. For instance, Bider et al. [8] show an example business process architecture model for a university where a same process (teaching and learning) belongs to both the primary and support process categories as seen from the perspective of the objects handled in the process. Though not shown in their meta-model, Malinova and Mendling [6] recognize the use of sub-categories, providing the example of local and global support processes, where the local processes support only a subset of the organization's core processes.

A *Phase* is defined as a cluster of processes that are executed at the same time. The temporal aspects of phase are not further defined. But the idea is clear and can for instance be observed in the top 'milestones' level of the framework for business process model decomposition proposed by [17], where the process map structures the most important processes in phases that end in milestones. Outside the BPM world, we see a similar clustering of project management processes in the PMBOK framework for project management [18] according to whether they are initiating, planning, executing, controlling or closing processes, i.e., referring to broad phases in a project's lifecycle.

Categories and phases are just two possible ways of grouping business processes. As shown in the concept map of Figure 4, belonging to a category or phase are properties of business processes that can be used as criteria to group processes that have a similar purpose or are executed concurrently. But any other property that processes have in common can be used as grouping criterion, including enterprise architecture elements like services offered, resources used/consumed, responsible actors and objects attained. The reviewed literature includes a great variety of process properties that can be used as grouping criterion: goals [1, 9], business functions [1, 9], key performance indicators [9]; locations [9]; organizational units [5]; access channels [2, 5]; product life cycle stages [5]; essential business entities [13]. In particular, concepts of goal and performance indicator are essential for when employing business process architecture for analysis and improvement of strategic alignment and performance of an organization's collection of business processes.

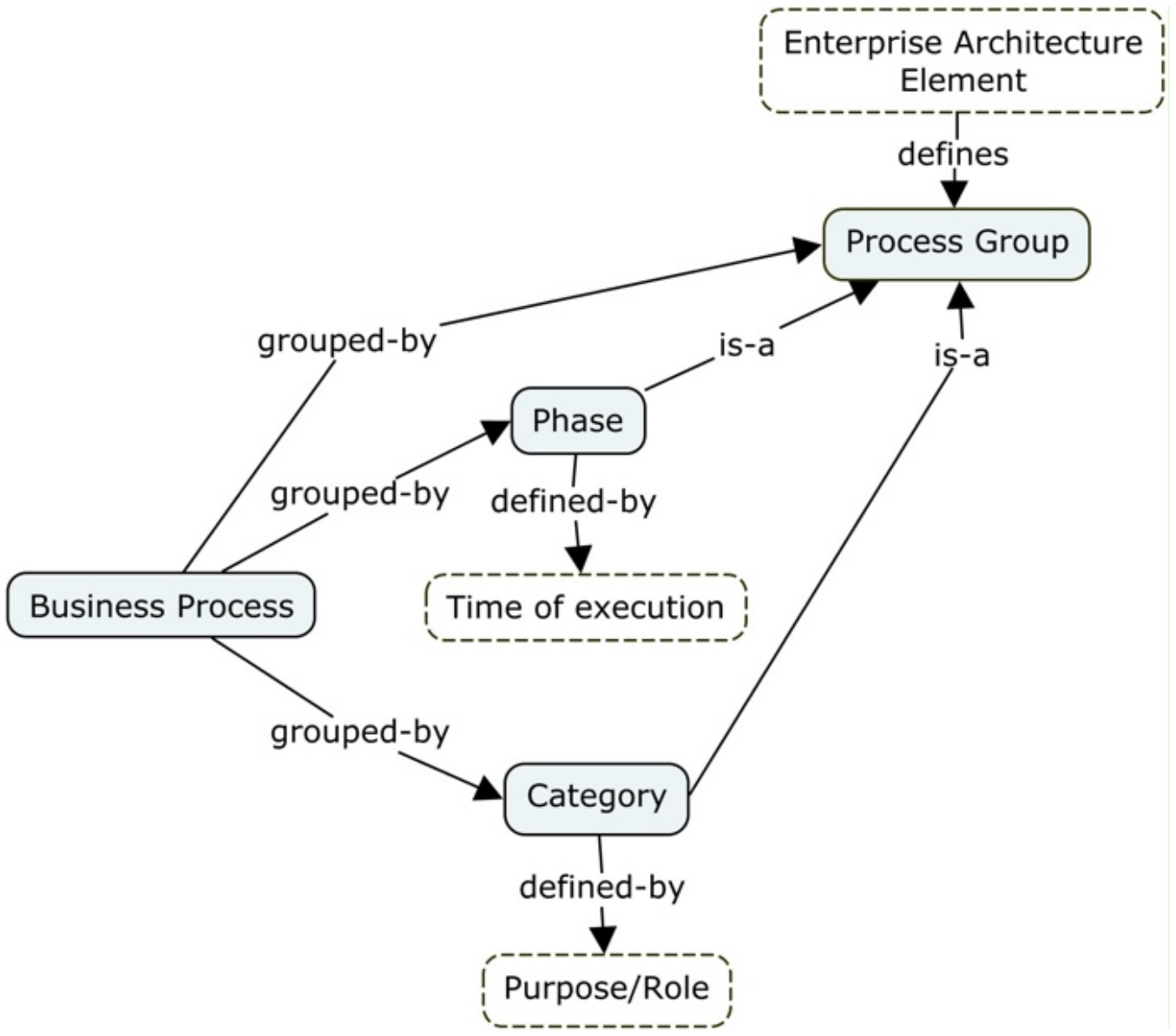

**Fig. 4.** Grouping business processes in categories and phases.

## 8. Business Process Relations

Finally, we review the different ways in which the processes shown on a process map can be related. Our BPA literature review identified four broad types of *Relations*: ordering, grouping, decomposition and specialization. We already discussed *Grouping* which does not relate processes amongst themselves but relates them to a *Process Group* based on defined grouping criteria like belonging to a category or phase or having some other property in common like the relationship to a particular enterprise architecture element (see Figure 4).

Different types of ordering relations have been identified. A *Trigger* relation between two processes indicates that one process triggers the execution of another process. The concept map depicted in Figure 2 shows that triggers are associated with internal or external inputs (e.g., execution requests, notifications, case objects). *Flow* means that a process passes information to another process but without triggering the start of that process. The executional semantics of trigger and flow has been formalized by Eid-Sabbagh, Dijkman and Weske [19] using patterns and anti-patterns, though their proposal necessitates the modelling of start, end and intermediate events of business processes (i.e., a process is no longer seen as an atomic unit of the process map, which violates Req. 1). Nevertheless, their formalization further clarifies the distinction between trigger and flow. Trigger means that an instance of one process instantiates and starts the execution of a second process. Flow means that an instance of one process passes information to an instance of another process that has already started, hence a flow relation can never instantiate and start a process. Whether a process instantiates and starts another process or not, both trigger and flow relations imply some sequential ordering between the processes (see Figure 5). An already started process acts and then a subsequent process is started and/or passed on some information, hence the prior process always needs to do something first, which is followed up by the subsequent process.

Apart from ordering relations, processes can be related through decomposition and specialization. According to Malinova and Mendling [6], *Decomposition* is a one-to-many relation that relates a superordinate process to one or more sub-processes. The situation in which a sub-process can be shared amongst superordinate processes (e.g., call activities to reusable processes in BPMN) is not considered in [6]. We choose to decouple decomposition and trigger relations in our conceptualization of process map. Decomposition shows which processes are executed as sub-processes of other processes, using shared aggregation semantics (i.e., many-to-many and not existentially dependent), while sequential ordering is an orthogonal relational dimension between processes that shows the order in which processes are executed (see Figure 5). While a superordinate process triggers the execution of its sub-processes, these sub-processes can be related (or not) amongst each other with trigger or flow relations and thus sequential ordering does not imply a decomposition relation between processes.

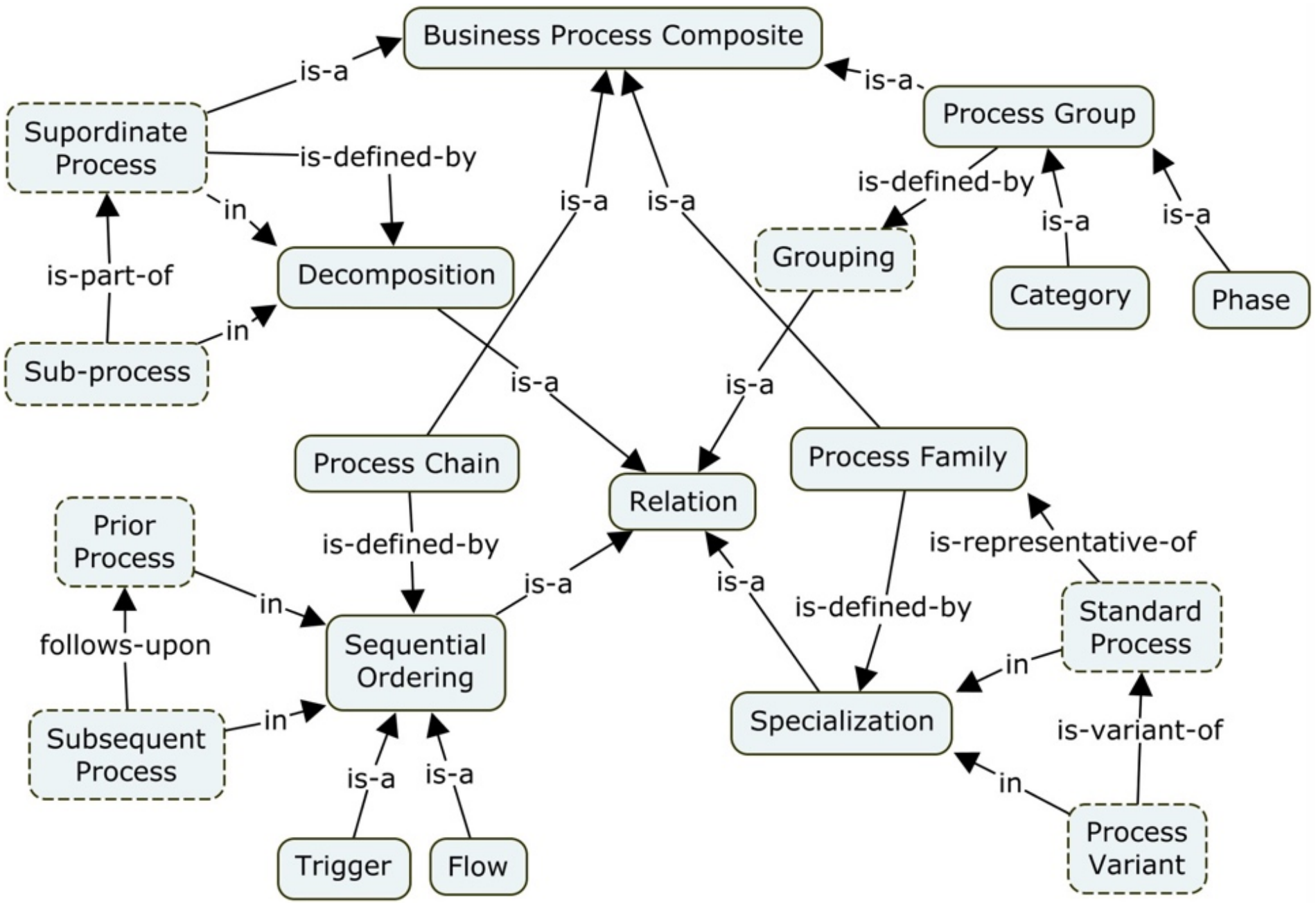

**Fig. 5.** Business process relations.

*Specialization* is used by Malinova and Mendling [6] to represent the relation between a standard process and its process variants. It is, however, not clear why they consider specialization also as a trigger relation. A specialization relation can be many-to-many. Van Nuffel and De Backer [9] give the example of an Executive Recruiting process variant of two main processes: HR Recruiting and HR Executive. The main or standard process is representative of the entire process family (see Figure 5).

Trigger, flow, decomposition and specialization are all covered in the meta-model of Malinova and Mendling [6], while grouping is implicitly so (e.g., through associating processes with categories and phases). In the reviewed BPA literature, we identify the need to also explicitly represent composites of related processes on the process map, like *Process Chains* (i.e., a whole of processes related through sequential ordering relations) and *Process Families* (i.e., standard processes representing not only themselves, but also their possible variants). For instance, the business process decomposition of Van Nuffel and De Backer [9] is specifically meant to represent process families.

Hence, we discover a need to not only represent business processes on a process map, but also abstractions of higher-order structures like process groups, process families and process chains. We formulate this need as a third solution requirement consistent with our research goal of designing a general representation language for process maps within the context of our architecture description conceptual frame of reference:

***Req. 3: It should be possible to show on a process map composites of business processes that result from the application of different types of process relations.***

The following types of process relations have been discovered in the reviewed literature, and need to be supported:

- *(Req. 3a) Sequencing relations*: The execution of business processes may be ordered in time meaning that the execution of a first process is followed by the execution of a second process. Such ordering relations typically indicate that processes are steps in a process chain that serves a higher-level goal. For instance, the requisition process and the purchasing process are steps of the Purchase-to-Pay (P2P) end-to-end process where requisition is performed before purchasing. Identifying ordering relations is important as changes applied to a prior process may affect the design and execution of a subsequent process.
- *(Req. 3b) Decomposition relations*: A business process can be a sub-process of another business process, like the sales order data entry process that is a sub-process of the sales order handling process. The steps of a process chain are sub-processes of the process chain. Decomposition can take the form of shared aggregation. For instance, a customer data verification process can be a sub-process of both a sales process and an after-sales service process. Decomposing business processes into sub-processes relates processes hierarchically which is important as BPM actions taken on sub-processes may affect their superordinate processes.

- *(Req. 3c) Grouping relations*: Business processes can be related through joint membership of a process group. From the moment business processes have something in common, a process group can be defined. For instance, a credit sales process and a cash sales process are members of the group of sales processes. Both processes share the goal of selling products or services to customers but differ in the manner in which customers pay for their sales. In principle, any property of processes can be used to form process groups. Defining process groups allows abstracting from certain differences between processes to see 'the forest through the trees', which can be important especially for organizations with large numbers of business processes.
- *(Req. 3d) Specialization relations*: A business process can be a specialized version of another business process, like the job student recruitment process that specializes the personnel recruitment process. A business process and its child processes form together a process family in which the child processes are process variants and the parent process becomes a standard process for these process variants. A process group, like the sales processes group, can be a process family, but is not necessarily so as there might be no standard sales process defined. Identifying specialization relations is important as changes applied to a parent process may have consequences for the child processes. Note that the implementation of specialization is not considered at the abstraction level of the business process architecture. One approach for instance is to define variation points in a standard process, which can be filled differently for the process variants [20].

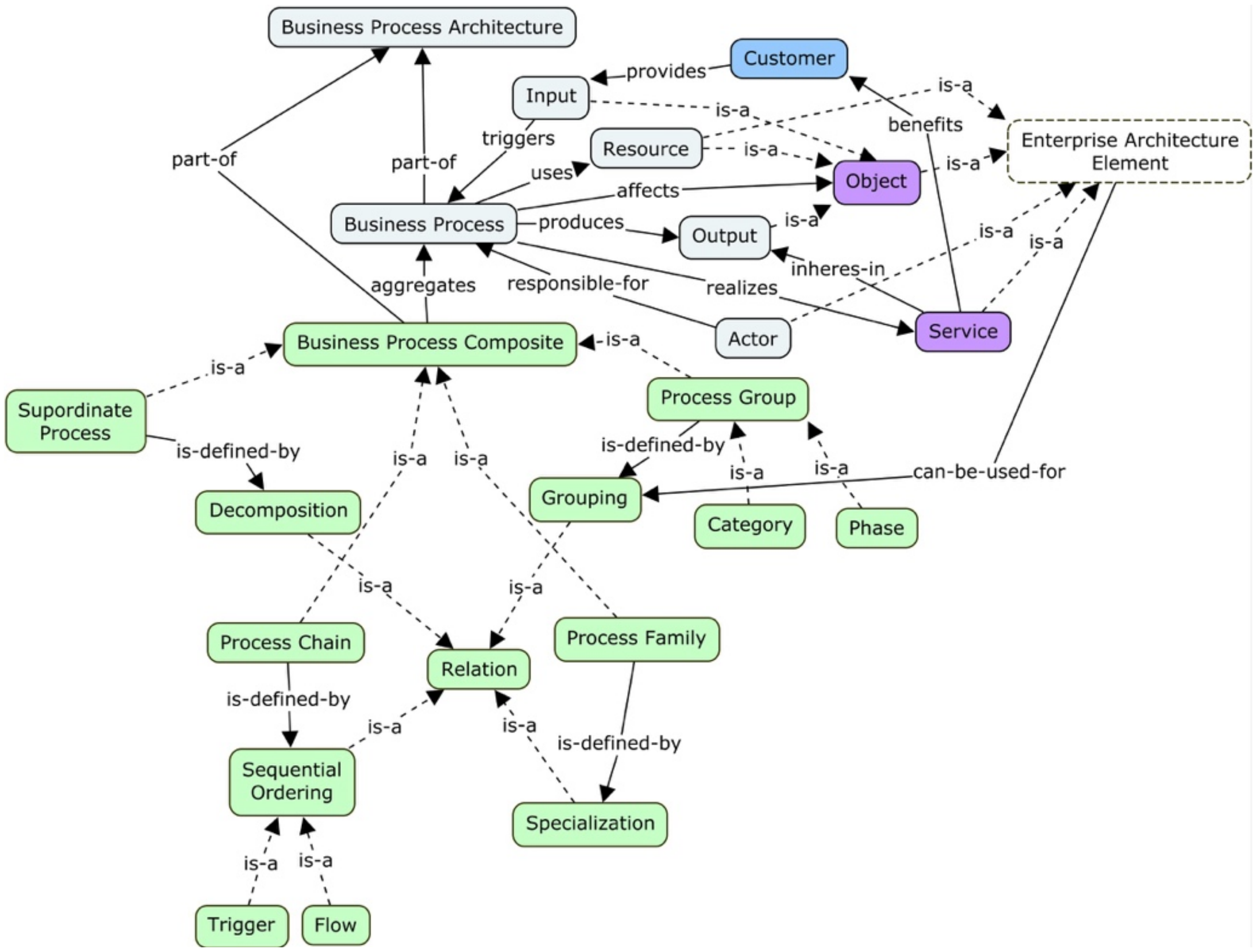

**Fig. 6.** Main process map concepts and relations.

## 9. Conclusion

In this report, we presented a new conceptualization of the process map as a business process architecture model that can be integrated into a broader enterprise architecture description. Figure 6 presents a concept map that shows the main concepts and relations of the proposed process map conceptualization.

Our research addresses the lack of 'architectural point of view' in current BPA research that aims at designing a modelling language for process maps as entry-level models of business process model hierarchies. The main requirements for a business process architecture description language that we derived from our conceptualization provide design principles for its meta-model: (i) a process map is a business process abstraction that provides a black-box view on the organization's business processes; (ii) the search for maximal integration with enterprise architecture description; and (iii) the recognition of different kinds of business process composites.

This report complements a research paper, entitled Architecting Business Process Maps, that reports on the design of such meta-model.

## Acknowledgements

This work is partially supported by GINSENG (TIN2015-70259), Ministerio de Economía, Industria y Competitividad (MINECO) y el Fondo Europeo de Desarrollo Regional (FEDER) and G3Soft (Model Engineering for Government and Management of Global Software Development, SBPLY/17/180501/000150), Consejería de Educación y Ciencia, Junta de Comunidades de Castilla-La Mancha y FEDER. Geert Poels was supported by a Ghent University Special Research Fund grant for sabbatical leave (Feb. 1, 2016 – Aug. 15, 2016).